\documentclass[5p]{elsarticle}  
\usepackage{natbib}
\usepackage{subfigure}
\usepackage{url}
\usepackage{graphicx}
\usepackage{float}
\usepackage{array}

\begin{document}

\begin{frontmatter}

\title{Baryon density extraction and isotropy analysis of Cosmic Microwave Background using Deep Learning}

\author[1]{Amit Mishra} 
\author[1]{Pranath Reddy}
\author[1]{Rahul Nigam}

\address[1]{Birla Institute of Technology \& Science Pilani - Hyderabad Campus, Hyderabad, India}

\begin{abstract}
The discovery of cosmic microwave background (CMB) was a paradigm shift in the study and fundamental understanding of the early universe and also the Big Bang phenomenon. Cosmic microwave background is one of the richest and intriguing sources of information available to cosmologists and one parameter of special interest is baryon density of the universe. Baryon density can be primarily estimated by analyzing CMB data or through the study of big bang nucleosynthesis(BBN). Hence, it is necessary that both of the results found through the two methods are in agreement with each other.  Although there are some well established statistical methods for the analysis of CMB to estimate baryon density, here we explore the use of deep learning in this respect. We correlate the baryon density obtained from the power spectrum of simulated CMB temperature maps with the corresponding map image and form the dataset for training the neural network model. We analyze the accuracy with which the model is able to predict the results from a relatively abstract dataset considering the fact that CMB is a Gaussian random field. CMB is anisotropic due to temperature fluctuations at small scales but on a larger scale CMB is considered isotropic, here we analyze the isotropy of CMB by training the model with CMB maps centered at different galactic coordinates and compare the predictions of neural network models.
\end{abstract}

\end{frontmatter}


\label{sec:intro}
\section{Introduction}
CMB is an electromagnetic radiation whose wavelength lies in the microwave region of the spectrum. The CMB maps display the temperature fluctuations of the photons that originate from a very early time when the universe was about one-millionth of its present size. These fluctuations follow an almost random gaussian distribution(Bucher, M., 2015) which extends onto the pixel intensities of the digital images of  mollweide projections of these maps. Since there is no spatial coherence between the CMB maps considering they are gaussian random fields, it is rather an intriguing task to check the accuracy with which machine learning models are able to correlate the gaussian maps to cosmological parameters considering the abstract nature of the dataset. Our objective is to build and study the ability of a machine learning model to predict baryon density of a given CMB map and also analyse the isotropy of CMB by comparing the results of models trained with maps centred at different galactic coordinates. Although there is a loss of information while training the model using rendered images of CMB, we have taken only a single quantity (baryon density) as our parameter of interest with a sole purpose of demonstrating the ability of a machine learning model to extract useful information from CMB data.

The variations in temperature of Cosmic Microwave Background (CMB) are similar to the ripples on the cosmic pond and enclose a lot of information about the universe. To collect this information we look at the scales at which these temperature fluctuations occur.  The amount of temperature fluctuations (in micro Kelvin) is plotted against the multipole moment (\textit{l}). This is the angular power spectrum graph of a CMB temperature map as shown in figure 1. Such graphs contain a number of peaks which provide us with a lot of information and we exploit this for our use.

The first peak is an indication of the geometry of the universe, whether it is flat or curved (Hu, Wayne, et al., 2004). CMB radiation is distorted by the curvature of the universe since the radiation comes from all directions of the visible universe. The fluctuations will appear undistorted if the universe is flat. The fluctuations would appear magnified if the universe is positively curved and de-magnified if it is negatively curved.

The second peak reveals information about the amount of baryon present in the universe. Due to the initial fluctuations in the universe, all matter would tend to gravitationally group towards the higher density fluctuations. However, baryon matter which is interactive with light would heat up as it clumps up, and the resultant pressure would try to push against the grouped matter (Hu, Wayne, et al., 2004). This implies that the second peak will be more damped if there is more matter. \\
Thus, the ratio of the first and second peak gives us the baryon density which we use for each map for training our model.

\begin{figure}[H]
	\centering
	\includegraphics[width = 2.5in]{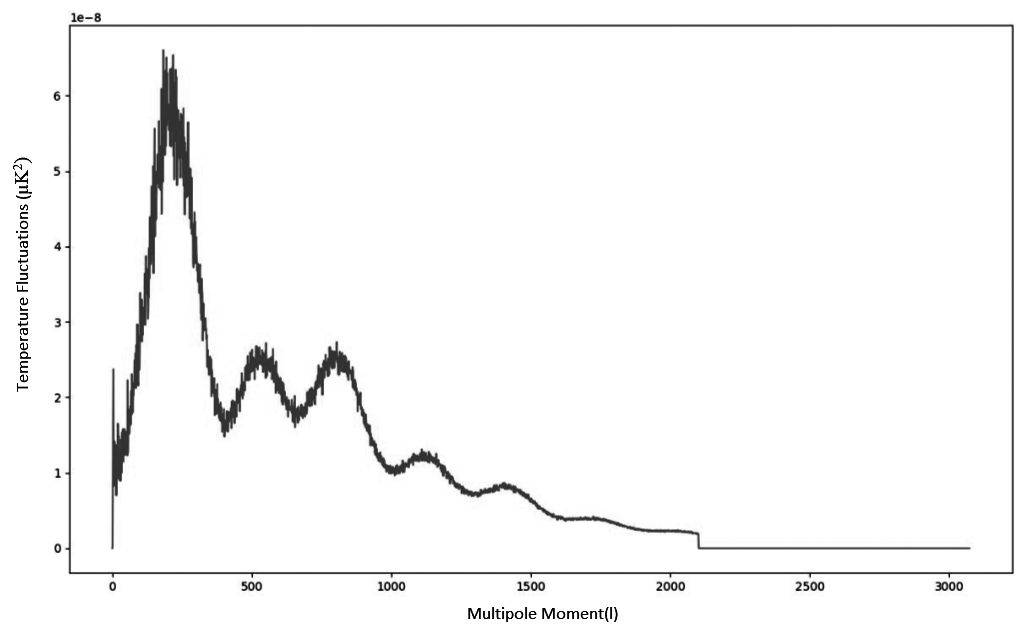}
	\caption{The angular power spectrum of a random CMB temperature sky map}		 	
\end{figure}
The anisotropies of CMB is determined by two factors, namely, acoustic oscillations and diffusion damping. The pressure of the photons tends to remove anisotropies, whereas the gravitational attraction of matter, makes them collapse to form over-densities. These two effects conflict and compensate each other to create acoustic oscillations, which gives CMB its characteristic peak structure. The resonant frequency at which photons decouple at a certain mode at its maximum amplitude corresponds to the peaks in the power spectrum of CMB.
The temperature anisotropy at any point on the sky ($\theta$,$\phi$) can be expressed in spherical harmonics as: \\
\begin{equation}
\frac{\Delta T}{T}(\theta ,\phi ) = \sum_{lm}^{ }a_{lm}Y_{lm}(\theta,\phi)
\end{equation}

The temperature anisotropies of the CMB are believed to be a result of the inhomogeneous matter distribution at the time of recombination. Since Compton scattering is an isotropic process, any primordial anisotropies should have been erased before decoupling. This provides proof to the results that are derived from the anisotropy that is observed as the result of density perturbations that facilitated the formation of galaxies and clusters. Anisotropy in temperature provides concrete evidence that such inhomogeneity in density existed in the early universe, supposedly in the scalar field of inflation caused by quantum fluctuations or through topological defects resulting from a phase transition.

Neural networks have been increasingly in use to approach and solve physics problems, especially in cosmology.  Models like variational autoencoders has been used for CMB image inpainting to reduce the uncertainty of parametric estimation (Yi, K., Guo, Y., Fan, Y., et al. 2020). Deep convolutional neural network or ResUNet has been used for lensing reconstruction of CMB (Caldeira et al., 2019). Removing foreground components from CMB has been achieved using a Bayesian spherical convolutional neural network that captures both spectral and morphological aspects of the foregrounds (Petroff, M. et al., 2020). Continuing this trend, in this paper we leverage neural networks for estimating baryon density of a given CMB temperature distribution and also verify its isotropic nature. 

\section{Methodology}
We use \textit{Code for Anisotropies in the Microwave Background or CAMB} to generate the CMB temperature angular power spectra data. CAMB is a cosmology library used to calculate CMB, lensing, source count and dark-age 21cm angular power spectra (Lewis A., Challinor A., 2014). CAMB takes several parameters as input to generate a \textit{FITS} file containing the initial angular power spectrum data of the universe. The Curved correlation function is used as the lensing method and we include reionization. Other physical parameters which are input to CAMB include Hubble constant, the temperature of CMB(2.7255 Kelvin), baryon density(0.0226), cold dark matter density(0.112), the effective mass density of dark energy, maximum multipoles data, redshift(11), helium fraction(0.24) and so on(\textit{LAMBDA-Tools}, NASA, 2014). The power spectrum file generated is used by standard cosmological analysis python package \textit{healpy} (Gorski, K. M., Hivon, E., Banday, A. J., et al. 2005) which is used to handle pixelated data on a sphere, to generate random gaussian CMB temperature maps. 2350 such maps are created. Figure 2 is one such CMB temperature map created using healpy. Anisotropy from dipole effect due to the movement of the earth relative to CMB rest frame and galactic contaminants along the equator corresponding to the galactic plane is removed while generating the temperature maps (Bucher, M., 2015)(Planck Collaboration, 2011). The generated full sky maps have the galactic center at the center of the mollweide projection.

Also, baryon density(extracted from the power spectrum) corresponding to each of the generated temperature maps is stored in a separate \textit{CSV} file.

We then snip 64x64 pixel(corresponding to 28x28 deg\textsuperscript{2}) size images from the whole sky map using OpenCV and store them separately(figure 3). The images are snipped only along the equator such that these portions could be treated as flat. In addition, we also rotate the CMB whole sky map along 4 different axis , moving the center of the maps to different galactic coordinates and repeat the same process in order to verify isotropy at large scales. These images will be treated as our input into the training model. We end up with 141,250 cropped images to train our model.
\begin{figure}[H]
	 
	\centering
	\includegraphics[width = 2.3in]{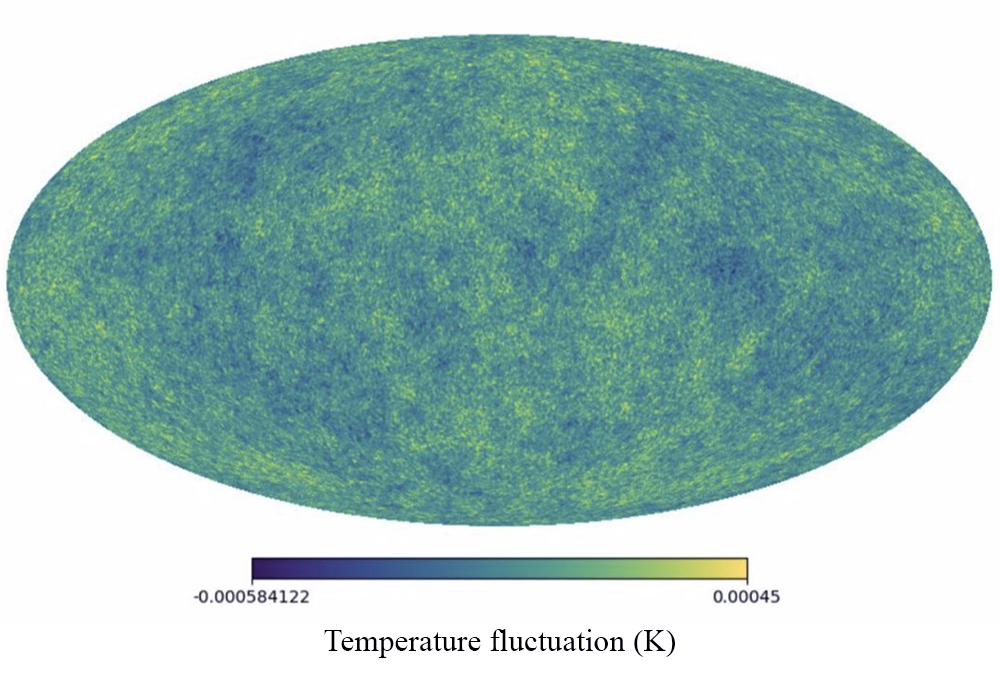}
	\caption{A random full sky CMB temperature map generated using healpy and CAMB}		 	
\end{figure}
\begin{figure}[H]
	 
	\centering
	\includegraphics[width = 3in]{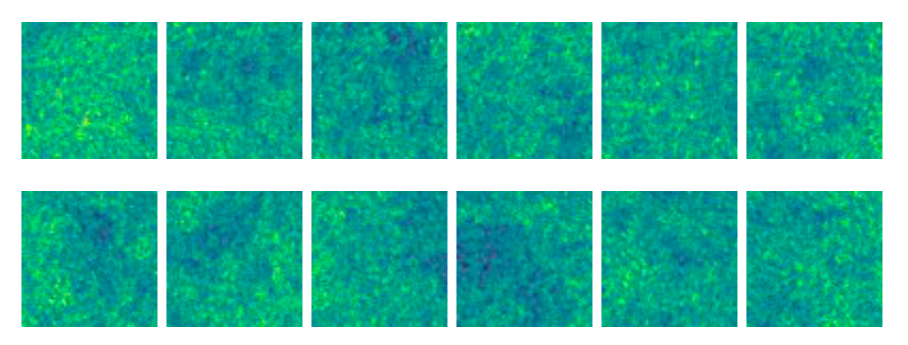}
	\caption{Sample patches cropped along the equator of the full sky CMB temperature map}		 	
\end{figure}

To remove outliers from our data, we have gone with the box plot approach and found the upper quartile (\textit{Q3}), lower quartile (\textit{Q1}), and also the inter-quartile distance (\textit{IQR = Q3 - Q1}) of all the baryon density values in our dataset and then removed all the data points that are greater than (\textit{Q3 + 1.5*IQR}) or lower than (\textit{Q1 -1.5*IQR}).

Furthermore, to aid our model in better extracting the features, we plot the pixel intensities of the grayscale training images and fit a Gaussian to the plot as shown in figure 4, and then replace the pixels with intensities that fall below a probability value of 0.05 with the average values of the remaining pixels to remove the outliers of an individual CMB distribution. Figure 5 shows a sample CMB distribution and a map with pixels with lower than set threshold intensity set to 0 to demonstrate the outliers .For additional reference, the data along the galactic coordinate (0\textsuperscript{0},0\textsuperscript{0}) has a range [0.026019, 0.049838] and a mean and standard deviation of 0.039272 and 0.00192502. The values of input CMB images are scaled to the range [0,1] before feeding them to the network.

\begin{figure}[H]
	 
	\centering
	\includegraphics[width = 3in]{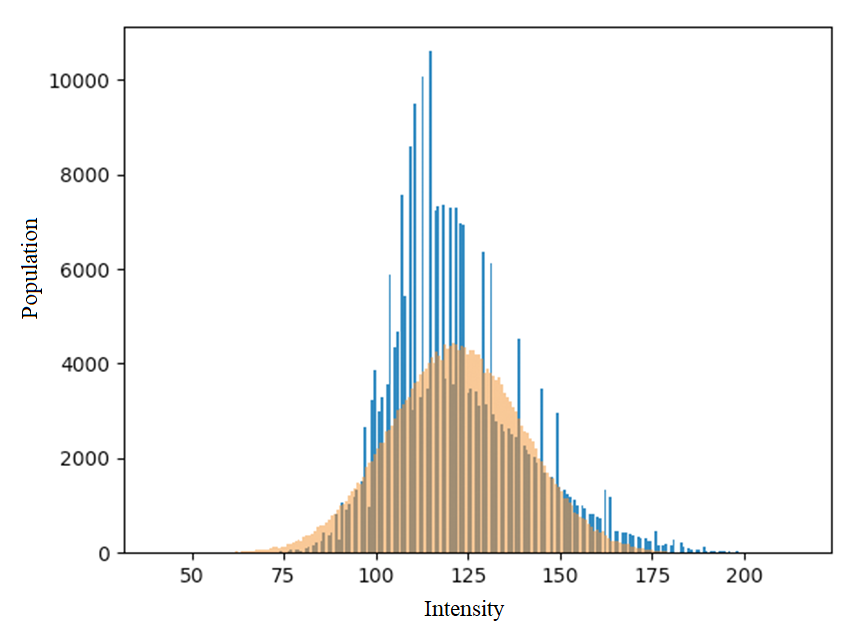}
	\caption{Fitting the pixel intensities with a Gaussian distribution(Population vs Pixel Intensities)}		 	
\end{figure}

\begin{figure}[H]
  \centering
  \begin{tabular}{@{}c@{}}
    \includegraphics[width = 2.5in]{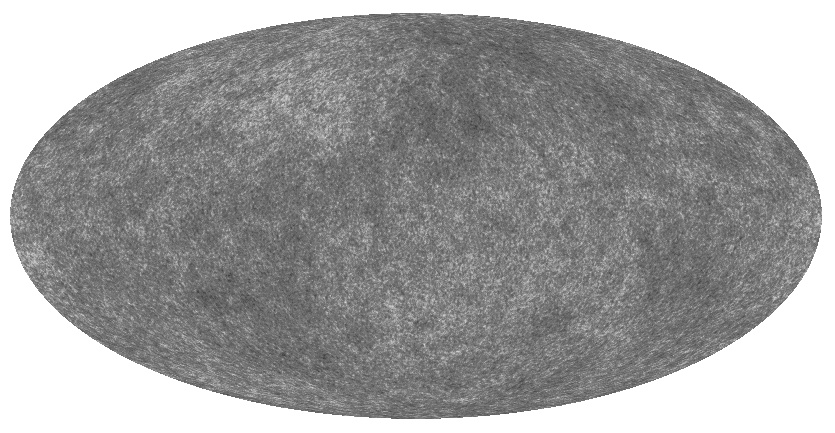} \\[\abovecaptionskip]
    \small (a)
  \end{tabular}

  \vspace{\floatsep}

  \begin{tabular}{@{}c@{}}
    \includegraphics[width = 2.5in]{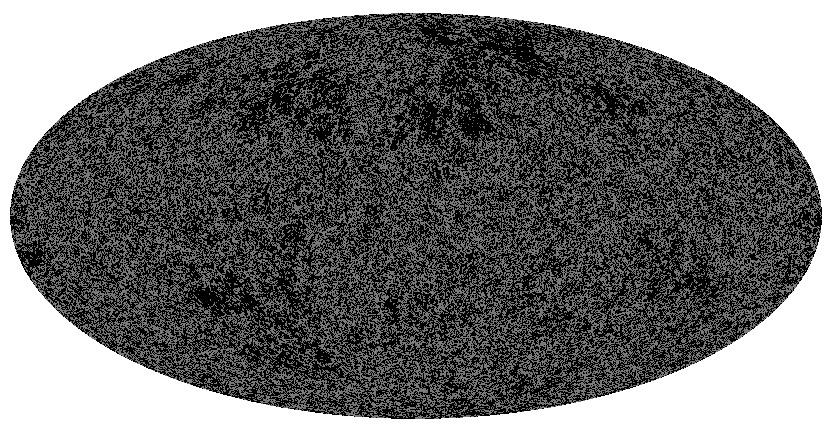} \\[\abovecaptionskip]
    \small (b) 
  \end{tabular}

  \caption{(a) Grayscale image of a test CMB temperature map and (b) the same map with pixels with lower than set threshold intensity deleted to demonstrate the outliers.}
\end{figure}

\section{Analysis}
To estimate the baryon density of a given CMB distribution, we will be using a CNN and an MLP based regression model. Convolutional neural networks (CNN) are one of the most famous sets of neural network architectures used for classifying images due to their ability to learn non-linear hierarchical structures and lower computational costs. CNN takes advantage of local spatial coherence of the input (Rippel, Snoek, Adams, 2015) because we assume that the spatially close images used for training are correlated, but in the case of the CMB dataset, the pixels in the images are random noise following a gaussian distribution. Hence It will be an interesting exercise to observe how the CNN model performs in our regression problem. The architecture of the CNN model is shown in figure 6.

\begin{figure}[H]
	 
	\centering
	\includegraphics[width = 1.7in]{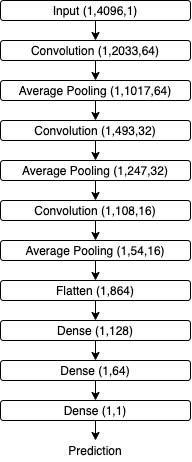}
	\caption{Architecture of the CNN model}		 	
\end{figure}

A multilayer perceptron is one of the most commonly used architectures of feedforward artificial neural networks. A multilayer perceptron comprises of three classes of layers and nodes, the input layer, hidden layers , and an output layer. Each node in a layer is connected to the nodes of the next layer via a non-linear activation function. Multilayer perceptron makes use of one of the most famous techniques of supervised learning called backpropagation (Goodfellow, Bengio, Courville, 2018) for training the network. A multilayer perceptron can be distinguished from a linear perceptron from its characteristic use of fully connected multilayers. This makes multilayer perceptrons suitable for working with non-linearly separable data (Bullinaria, 2015).
Multilayer perceptrons are often informally referred to as vanilla networks (Hastie, Trevor, et al., 2017).
A multilayer perceptron can be perceived  as a logistic regression classifier. The input is transfigured with the help of a learned non-linear transformation. The intermediate layers are often mentioned as a hidden layer. These hidden layers make the multilayer perceptron a “universal approximator” (Sifaoui, Abdelkrim, Benrejeb, 2008). The architecture of the MLP model is shown in figure 7.

\begin{figure}[H]
	 
	\centering
	\includegraphics[width = 1.7in]{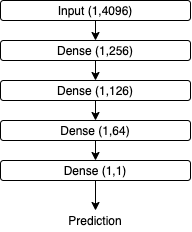}
	\caption{Architecture of the MLP model}		 	
\end{figure}

The weights of the fully connected layers are updated once a batch of data has been passed through the network by measuring the error of the output with the expected result (predetermined labels), this is the essence of learning in neural networks and is carried out with the help of an iterative algorithm called backpropagation. This is an example of supervised learning. Backpropagation uses an iterative optimization algorithm called gradient descent (Goodfellow, Bengio, Courville, 2018) to update the weights of the network. 

The configuration of the MLP network is tabulated in Table 1.

\begin{table}[H]
\small
\caption{MLP Configuration}
\begin{tabular}{| >{\centering\arraybackslash}m{1in} | >{\centering\arraybackslash}m{1in} | >{\centering\arraybackslash}m{1in} |}
\hline
No of hidden layers & No of Nodes in each hidden layer & Learning Rate\\
\hline
3 & 256,128,64 & 0.0001\\
\hline
\end{tabular}

\begin{tabular}{| >{\centering\arraybackslash}m{1in} | >{\centering\arraybackslash}m{1in} | >{\centering\arraybackslash}m{1in}|}
\hline
Batch size & No of epochs & Regularization parameter\\
\hline
512 & 5000 & 0.01\\
\hline
\end{tabular}
\end{table}

The learning rate determines how fast the weights or the coefficients of the network are updated. An epoch can be defined as the number of times the algorithm perceives the entire data-set. Hence, an epoch is completed when all the samples of the data have been perused. An iteration can be defined as the number of times a “batch of data” has been passed through the algorithm. In the case of a multilayer perceptron, that means the forward pass and backward pass. Hence, an iteration is completed once a batch of data has passed through the network. The batch size is the number of training examples passed through the network at once (Shen, 2017 \& Svozil, Kvasnicka, Pospichal, 1997).

We have used the tensorflow and keras libraries to implement our CNN and MLP models. We have used Adam optimization algorithm (Kingma, Ba, 2017) instead of the traditional stochastic gradient descent for updating the weights of the network (Michelucci, Umberto, 2018). 

We have used L2 regularization, also known as ridge regularization to prevent our model from overfitting. In L2 regularization, we add a squared error term as a penalty to the loss function (Goodfellow, Bengio, Courville, 2018). 

The training of the network is done in the Google Cloud platform using a Tesla K80 GPU.

\section{Results}

\begin{figure}[H]
	 
	\centering
	\includegraphics[width = 2.5in]{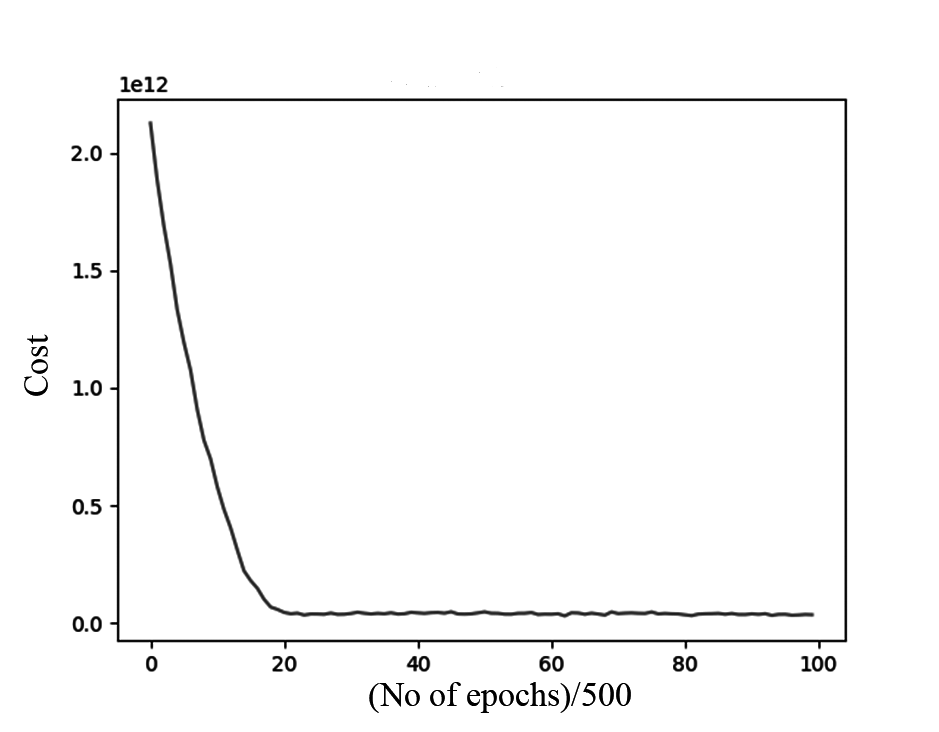}
	\caption{The change in training cost of models ( vs no of (epochs/500) ) trained on data corresponding to a specific galactic latitude and longitude (-30\textsuperscript{0},90\textsuperscript{0}) combination at the center of the mollweide projection ( input images are cropped along the central horizontal line of the projection ), shows the general trend of change in cost.}		 	
\end{figure}

We have used Hold-out cross validation with an 80-20 split for evaluating the performance of our models. And for the metrics, we have gone with Mean Absolute error (MAE), Root Mean squared error (RMSE), Mean Magnitude of the Relative Error (MMRE), R2 Score and Pearson Correlation. R2 Score and Pearson Correlation are commonly used statistical measures that compute the closeness between the real labels and the regression model's predicted values.
\\

\begin{equation}
MAE = \frac{1}{n}\sum|y-yp|
\end{equation}

\begin{equation}
RMSE = \sqrt{\frac{1}{n}\sum|y-yp|^2}
\end{equation}

\begin{equation}
MMRE = \frac{1}{n}\sum\frac{|y-yp|}{y}
\end{equation}
where,\\
		\textit{ n: no of samples}\\
		\textit{y: actual value}\\
	        \textit{yp: predicted value}

The CNN and MLP models are trained and tested taking five different galactic coordinates at the center of the mollweide projection and cropping images along the central horizontal line to test the isotropy of the CMB distribution. The loss convergence of a CNN model trained along the galactic coordinates (-30\textsuperscript{0},90\textsuperscript{0}) is shown in figure 8. The results of the CNN model are tabulated in Table 2, and the results of the MLP model are tabulated in Table 3. We have also shown a scatter plot of predictions along the galactic coordinates (0\textsuperscript{0},0\textsuperscript{0}) in figure 9.

Furthermore, to study the isotropy of our CMB samples, we have also presented the Relative error between different CNN models trained at different galactic coordinates in Table 4.

\begin{figure}[H]
	 
	\centering
	\includegraphics[width = 3in]{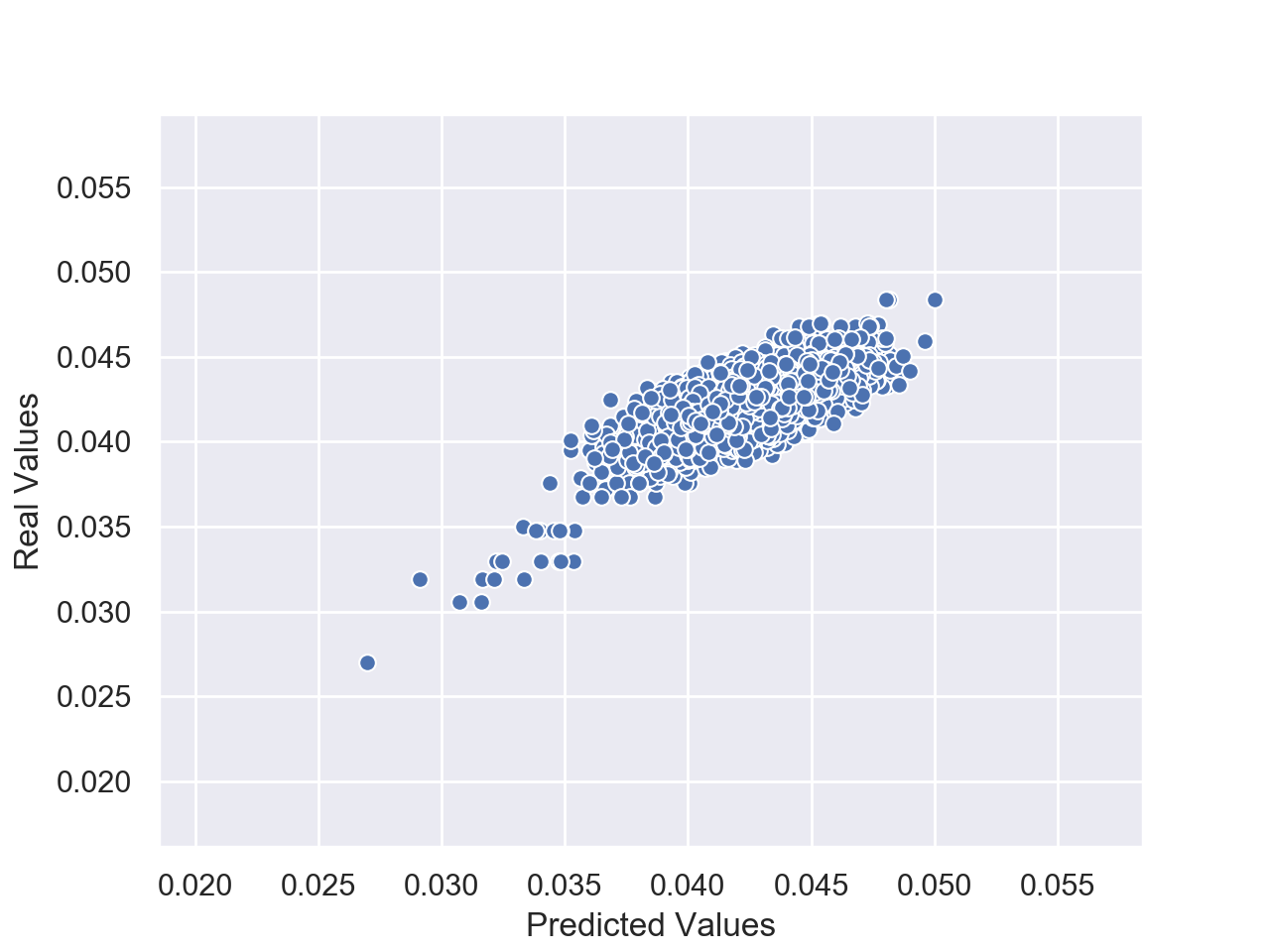}
	\caption{A scatter plot showing the predicted bayon density values for a model trained on the data corresponding to the galactic coordinates (0\textsuperscript{0},0\textsuperscript{0})}		 	
\end{figure}

\begin{table}[!h]
\small
\caption{Results for CNN models trained with different galactic coordinates at center}
\begin{tabular}{| >{\centering\arraybackslash}m{0.3in} | >{\centering\arraybackslash}m{0.8in} | >{\centering\arraybackslash}m{0.8in} | >{\centering\arraybackslash}m{0.8in} |}
\hline
Sl.No. & Galactic coordinates & MAE & RMSE
\\
\hline
1	&(0\textsuperscript{0},0\textsuperscript{0})	&0.0012148	&0.0015857   \\
2	&(30\textsuperscript{0},90\textsuperscript{0}) 	&0.0012985	&0.0016533   \\
3	&(-30\textsuperscript{0},90\textsuperscript{0}) &0.0012469	&0.0015921   \\
4	&(60\textsuperscript{0},90\textsuperscript{0})	&0.0012675	&0.0016243   \\
5	&(-60\textsuperscript{0},90\textsuperscript{0})	&0.0012889	&0.0016311   \\

\hline
\end{tabular}

\begin{tabular}{| >{\centering\arraybackslash}m{0.3in} | >{\centering\arraybackslash}m{0.8in} | >{\centering\arraybackslash}m{0.8in} |}
\hline
Sl.No. & Galactic coordinates & MRE
\\
\hline
1	&(0\textsuperscript{0},0\textsuperscript{0})	&0.026331\\
2	&(30\textsuperscript{0},90\textsuperscript{0}) 	&0.030122\\
3	&(-30\textsuperscript{0},90\textsuperscript{0}) &0.028343\\
4	&(60\textsuperscript{0},90\textsuperscript{0})	&0.029109\\
5	&(-60\textsuperscript{0},90\textsuperscript{0}) &0.029897\\

\hline
\end{tabular}

\begin{tabular}{| >{\centering\arraybackslash}m{0.3in} | >{\centering\arraybackslash}m{0.8in} | >{\centering\arraybackslash}m{0.8in} | >{\centering\arraybackslash}m{0.8in} |}
\hline
Sl.No. & Galactic coordinates & R2 Score & Pearson Correlation
\\
\hline
1	&(0\textsuperscript{0},0\textsuperscript{0})	&0.226719 &0.754609\\
2	&(30\textsuperscript{0},90\textsuperscript{0}) 	&0.209762 &0.745292\\
3	&(-30\textsuperscript{0},90\textsuperscript{0}) &0.219951 &0.752798\\
4	&(60\textsuperscript{0},90\textsuperscript{0})	&0.216103 &0.751073\\
5	&(-60\textsuperscript{0},90\textsuperscript{0}) &0.212899 &0.749151\\

\hline
\end{tabular}
\end{table}

\begin{table}[!h]
\small
\caption{Results for MLP models trained with different galactic coordinates at center}
\begin{tabular}{| >{\centering\arraybackslash}m{0.3in} | >{\centering\arraybackslash}m{0.8in} | >{\centering\arraybackslash}m{0.8in} | >{\centering\arraybackslash}m{0.8in} |}
\hline
Sl.No. & Galactic coordinates & MAE & RMSE
\\
\hline
1	&(0\textsuperscript{0},0\textsuperscript{0})	&0.0013289	&0.0016519   \\
2	&(30\textsuperscript{0},90\textsuperscript{0}) 	&0.0013879	&0.0016972   \\
3	&(-30\textsuperscript{0},90\textsuperscript{0}) &0.0013511	&0.0016648   \\
4	&(60\textsuperscript{0},90\textsuperscript{0})	&0.0013692	&0.0016698   \\
5	&(-60\textsuperscript{0},90\textsuperscript{0})	&0.0013773	&0.0016823   \\

\hline
\end{tabular}

\begin{tabular}{| >{\centering\arraybackslash}m{0.3in} | >{\centering\arraybackslash}m{0.8in} | >{\centering\arraybackslash}m{0.8in} |}
\hline
Sl.No. & Galactic coordinates & MRE
\\
\hline
1	&(0\textsuperscript{0},0\textsuperscript{0})	&0.031892\\
2	&(30\textsuperscript{0},90\textsuperscript{0}) 	&0.035289\\
3	&(-30\textsuperscript{0},90\textsuperscript{0}) &0.032723\\
4	&(60\textsuperscript{0},90\textsuperscript{0})	&0.033801\\
5	&(-60\textsuperscript{0},90\textsuperscript{0}) &0.035067\\

\hline
\end{tabular}

\begin{tabular}{| >{\centering\arraybackslash}m{0.3in} | >{\centering\arraybackslash}m{0.8in} | >{\centering\arraybackslash}m{0.8in} | >{\centering\arraybackslash}m{0.8in} |}
\hline
Sl.No. & Galactic coordinates & R2 Score & Pearson Correlation
\\
\hline
1	&(0\textsuperscript{0},0\textsuperscript{0})	&0.092853 &0.725313\\
2	&(30\textsuperscript{0},90\textsuperscript{0}) 	&0.089492 &0.709953\\
3	&(-30\textsuperscript{0},90\textsuperscript{0}) &0.091607 &0.721934\\
4	&(60\textsuperscript{0},90\textsuperscript{0})	&0.091238 &0.720486\\
5	&(-60\textsuperscript{0},90\textsuperscript{0}) &0.090379 &0.716433\\

\hline
\end{tabular}
\end{table}

\begin{table}[H]
\small
\caption{Relative error between different models trained with different galactic coordinates at center}
\begin{tabular}{| >{\centering\arraybackslash}m{1in} | >{\centering\arraybackslash}m{0.6in} | >{\centering\arraybackslash}m{0.6in} | >{\centering\arraybackslash}m{0.6in} |}
\hline
Galactic coordinates	&(0\textsuperscript{0},0\textsuperscript{0})	&(30\textsuperscript{0},90\textsuperscript{0})	&(-30\textsuperscript{0},90\textsuperscript{0})
\\
\hline
(0\textsuperscript{0},0\textsuperscript{0})		&-			&0.029728	&0.030671\\
(30\textsuperscript{0},90\textsuperscript{0}) 	&0.029728	&-			&0.034365	\\
(-30\textsuperscript{0},90\textsuperscript{0}) 	&0.030671	&0.034365	&-\\
(60\textsuperscript{0},90\textsuperscript{0})	&0.029723	&0.034801	&0.034580	\\
(-60\textsuperscript{0},90\textsuperscript{0})	&0.030752	&0.034343	&0.034401	\\
\hline
\end{tabular}

\begin{tabular}{| >{\centering\arraybackslash}m{1in} | >{\centering\arraybackslash}m{0.97in} | >{\centering\arraybackslash}m{1in} |}
\hline
Galactic coordinates	&(60\textsuperscript{0},90\textsuperscript{0})	&(-60\textsuperscript{0},90\textsuperscript{0})
\\
\hline
(0\textsuperscript{0},0\textsuperscript{0})		&0.029723	&0.030752\\
(30\textsuperscript{0},90\textsuperscript{0}) 	&0.034801	&0.034343\\
(-30\textsuperscript{0},90\textsuperscript{0}) 	&0.034580	&0.034401\\
(60\textsuperscript{0},90\textsuperscript{0})	&-		&0.034289\\
(-60\textsuperscript{0},90\textsuperscript{0})	&0.034289	&-		 \\
\hline
\end{tabular}

\end{table}

\section{Conclusion and future plans}
The application of deep learning methodologies on the CMB data has steered us to the verification of two well-established results related to the CMB using a completely different approach via deep learning. Firstly, we were able to predict baryon density with a satiable accuracy. The loss of accuracy can be credited to the fact that the CMB temperature maps are of very high resolution and we were bounded by limited computational power but most importantly the relatively random orientation of the pixels considering the fact that CMB is a Gaussian random field was a major complexity for the neural network model to extract any applicable features from the maps. Secondly, when training and predicting with models trained along different galactic latitude and longitude we were able to get very low error between the predictions which reaffirms the well known isotropic nature of CMB at larger scales. Although the training accuracy is low in the domain of precision physics, the test error values are impressive considering the fact that we were limited by the amount of data and also computational power of training our model. We are not trying to compete with traditional well established techniques with the power spectrum, but rather we are proposing a new domain for the study of CMB and subsequently as a demonstration we have chosen baryon density for the same. Although, in the case of extracting baryon density directly from temperature map there is a loss of information but there could be some property of CMB which could be found using deep learning but not from power spectrum. We hope to develop a deep neural network architecture tailored for CMB maps that is capable of correlating random noise to cosmological parameters.

\section{Data Availability}

To promote collaborative research, we have made the code used in this paper open source at

\textit{https://github.com/iamitm/cosmic} (The repository is yet to be properly organized and documented).

\end{document}